\documentclass[%
 reprint,
 superscriptaddress,
 amsmath,amssymb,
 aps,
 prl,
]{revtex4-2}

\usepackage{graphicx}  
\usepackage{dcolumn}   
\usepackage{bm}        
\usepackage[colorlinks=true, linkcolor=blue, citecolor=blue, urlcolor=blue]{hyperref}

\begin{document}

\title{Universal Configuration for Optimizing Complexity in Variational Distributed Quantum Circuits}

\author{J. Montes}
\email{jmontes.3@alumni.unav.es}
\affiliation{Grupo de Sistemas Complejos, ETSIME, Universidad Polit\'ecnica de Madrid, Rios Rosas 21, 28003 Madrid, Spain}

\author{F. Borondo}
\email{f.borondo@uam.es}
\affiliation{Departamento de Qu\'imica, Universidad Aut\'onoma de Madrid, Cantoblanco, 28049 Madrid, Spain}

\author{Gabriel G. Carlo}
\email{g.carlo@conicet.gov.ar}
\affiliation{Comisi\'on Nacional de Energ\'ia At\'omica, CONICET, Departamento de F\'isica, Av.~del Libertador 8250, 1429 Buenos Aires, Argentina}

\date{\today}

\begin{abstract}
Distributed quantum computing represents at present one of the most promising approaches to scaling quantum processors. Current implementations typically partition circuits into multiple cores, each composed of several qubits, with inter-core connectivity playing a central role in ensuring scalability. Identifying the optimal configuration—defined as the arrangement that maximizes circuit complexity with minimal depth—thus constitutes a fundamental design challenge. In this work, we demonstrate, both analytically and numerically, the existence of a universal optimal configuration for distributing single- and two-qubit gates across arbitrary inter-core communication topologies in variational distributed circuits. Our proof is based on a complexity measure based on Markov matrices, which quantifies the convergence rate toward the Haar measure, as introduced by Weinstein \emph{et al.}~\cite{Weinstein}. Finally, we validate our predictions through numerical comparisons with the well-established majorization criterion proposed in Ref.~\cite{MulticoreNuestro}.
\end{abstract}



\maketitle

\emph{Introduction.} 
Significant and very recent improvements in error-correction techniques~\cite{error,error2} give us hopes of reaching the large-scale fault tolerant quantum hardware in the not so distant future. In the meantime, the development of noisy intermediate-scale quantum (NISQ) processors~\cite{Preskill2018} has been friutful, allowing the first demonstrations of quantum supremacy~\cite{Arute}, for example. Experimental platforms~\cite{Madsen} and specialized applications~\cite{King} have become a tangible reality. In both cases, scaling devices to large numbers of qubits is necessary but demanding. Partitioning the processor into interconnected smaller modules gives rise to a distributed architecture which is a reasonable approach to tackle this problem (being analogous to classical multicore computing~\cite{multicore,Hetenyi2024}). Recently, a modular quantum processor equipped with a reconfigurable router enabling full connectivity has been demonstrated~\cite{Wu}, based on a mainboard linking multiple daughter boards and successfully implementing SWAP gates. This result adds to a series of notable achievements, including the coupling of processors through coaxial cables~\cite{Yam}, the interconnection of 35 photonic chips via optical devices culminating in the Aurora machine~\cite{Rad}, and the integration of two IBM processors to execute a circuit exceeding the capabilities of each individual device~\cite{IBM}.

In distributed quantum computing, the primary challenge lies in generating and maintaining high quantum complexity, which is essential for multicore devices to deliver practical utility. In most platforms, creating entanglement across distinct cores is substantially more demanding than generating correlations within each core~\cite{interconnects,interconnects2}. This technological limitation has motivated the development of circuit distribution strategies focused primarily on reducing total circuit depth and minimizing the number of inter-core quantum communication operations~\cite{dist1,dist2,dist3}. But optimizing these architectures is far from trivial and this has led to different strategies based on measuring the intercore qubit traffic for example~\cite{traffic}. 

At the same time, finding the minimal requirements that a given architecture must satisfy to ensure a sufficient level of complexity~\cite{work1} to support quantum advantage is crucial. To achieve this several characterization and benchmarking methodologies have been proposed. Among them, the recently formulated majorization criterion~\cite{work1} provides an efficient and practical tool to quantify the complexity of random circuits generated with different families of gates. It has proven highly useful both in quantum reservoir computing scenarios~\cite{qrc1,qrc2} and in assessing the complexity of current universal quantum processors~\cite{work2}.

In this work, we combine the majorization criterion~\cite{MulticoreNuestro} with a complementary framework based on Markov chain modeling in order to obtain the optimization of variationally distributed quantum circuits~\cite{Weinstein}. In particular, we construct a parametrized Markov matrix describing the evolution of the expected second moments of the density matrix expansion in the Pauli basis, and analyze its spectral gap as a function of the number of intracore steps and the topology of inter-core connections. In this way we show through numerical simulations and analytical arguments, we show that, independently of the specific architecture and the number of cores, there exists an optimal configuration: an intermediate number of local iterations preceding the inter-core entangling operations, which maximizes the spectral gap and minimizes the distance to the Haar measure. To our knowledge, this is the first study to establish analytically the conditions under which convergence toward highly complex states exhibits a maximum as a function of intracore depth. 

These findings not only provide a general principle to guide the design of distributed quantum circuits but also offer quantitative metrics that can inform the development of scalable quantum processors.

\emph{System definition and parametrization.}
We consider a distributed quantum circuit characterized by the number of cores, $N_c$. Each core contains $N_q$ qubits, yielding a total of $N_c \times N_q$ qubits in the system.

Rather than focusing on a specific wavefunction, we consider an ensemble of quantum states evolving under the action of randomly generated variational circuits. Our interest lies in the statistical evolution induced by the circuit layers, characterized by their average effect over the ensemble.

The circuit depth is determined by the sequence of gates applied in each layer. In every layer, we distinguish two types of operations: \emph{intracore} gates, which act exclusively within a core, and \emph{intercore} gates, which entangle qubits belonging to different cores.

Intracore gates are chosen to be universal within each core. Specifically, each intracore layer consists of $I$ gates:
\[
\hat{U}(I) = \hat{U}_1\,\hat{U}_2\,\dots\,\hat{U}_I.
\]
Let us denote by \( |\phi_{q,c}\rangle \) the state of the \(q\)-th qubit in core \(c\), and by \( |\psi\rangle \) the initial state of the entire system.

Their action on the state reads:
\[
\hat{U}(I)\,|\psi\rangle = \bigotimes_{c=1}^{N_c}\hat{U}(I)\left(\bigotimes_{q=1}^{N_q}|\phi_{q,c}\rangle\right).
\]
Each operator $\hat{U}_i$ is sampled as follows: with probability $p_1$, a single-qubit gate is applied to a qubit $j$ selected uniformly with probability $p_j = 1/N_q$; with probability $p_2 = 1 - p_1$, a two-qubit gate is selected. In this case, there are $(N_q-1)\,N_q$ possible pairs of qubits within the core, each chosen uniformly with probability $1/\bigl[(N_q-1)\,N_q\bigr]$.

After applying the intracore sequence, the intercore gates are incorporated. These are two-qubit operations acting on qubits located in different cores. The pairs of cores where such gates can be applied to define the \emph{topology} of the architecture.

In this work, we analyze four representative topologies: linear, ring, star, and fully connected. These correspond, respectively, to the following configurations: connections between adjacent cores; connections between consecutive cores plus the link between the first and the last; a central core connected to all others; and connections between all pairs of cores.

For each allowed link between two cores, a two-qubit gate is sampled uniformly over all possible qubit combinations participating in the operation. Since one qubit acts as control and the other as target, the total number of configurations is $2\,N_q^2$, and each is chosen with probability $1/(2\,N_q^2)$.

The combined action of intercore operations on the system state is:
\[
|\varphi\rangle = \hat{U}_{\mathrm{inter}}\,\hat{U}(I)\,|\psi\rangle =
\hat{U}_{\mathrm{inter}}\left(\bigotimes_{c=1}^{N_c}\hat{U}(I)\left(\bigotimes_{q=1}^{N_q}|\phi_{q,c}\rangle\right)\right).
\]
The resulting state \( |\varphi\rangle \) can be written as a superposition over the computational basis of \( N_c N_q \) qubits, 
$
|\varphi\rangle = \sum_{\gamma} \alpha_\gamma\,|\gamma\rangle,
$
where \( \alpha_\gamma \in \mathbb{C} \) and \( \{|\gamma\rangle\} \) denotes the standard computational basis.

This procedure defines a layer characterized by the parameters $N_q$, $N_c$, $I$, the topology, the types of single- and two-qubit gates, and the sampling probabilities for each intracore operation. Throughout this study, we consider single-qubit gates implemented as variational rotations with three independent angles drawn from the Haar measure, while two-qubit gates correspond to controlled-$Z$ (CZ) operations.

Having established the system definition, in the following we proceed to quantify its complexity as a function of these parameters.

\emph{Complexity quantification.}
To assess the distance between our ensemble of variational circuits and the ensemble of random states distributed according to the Haar measure, we employ two complementary approaches. The first is purely numerical and is based on the principle of \emph{majorization}.

Majorization consists in comparing two probability vectors after ordering their components in non-increasing order. Given two vectors $\mathbf{p}, \mathbf{q}\in\mathbb{R}^M$, we say that $\mathbf{p}$ is majorized by $\mathbf{q}$ (denoted $\mathbf{p}\prec\mathbf{q}$) if the following conditions hold:
\[
\sum_{i=1}^{k} p_i^{\downarrow}\le \sum_{i=1}^{k} q_i^{\downarrow}, \quad 1\le k<M,
\]
\[
\sum_{i=1}^{M} p_i = \sum_{i=1}^{M} q_i,
\]
where the superscript $\downarrow$ indicates decreasing ordering. This criterion serves as an indicator of uniformity, which we apply to the probability vectors obtained by measuring the output states of the circuits in the computational basis.

For each vector, we define the Lorenz curves given by the partial cumulants:
\[
\mathcal{F}_{\mathbf{p}}(k) = \sum_{i=1}^{k} p_i^{\downarrow},
\]
and analogously for $\mathcal{F}_{\mathbf{q}}(k)$. If $\mathbf{q}$ majorizes $\mathbf{p}$, the Lorenz curve of $\mathbf{q}$ lies above that of $\mathbf{p}$ for all $k$.

In our case, we compute the probability distributions resulting from applying the variational circuit $\hat{U}_{\mathrm{inter}}\hat{U}(I)$ repeated $L$ layers to the initial state
\[
|0\dots0\rangle = |0\rangle^{\otimes N_c N_q},
\]
and measuring in the computational basis. We denote these distributions as
\[
p_U(i) = \bigl|\langle i\,|\,(\hat{U}_{\mathrm{inter}}\hat{U}(I))^L\,|0\dots0\rangle\bigr|^2.
\]
From these, we obtain the cumulants $\mathcal{F}_{p_U}(k)$ for $k\in\{1,\dots,2^{N_c N_q}\}$ and compute their fluctuations:
\[
\mathrm{std}\bigl[\mathcal{F}_{p_U}(k)\bigr] = \sqrt{\langle \mathcal{F}_{p_U}(k)^2\rangle - \langle\mathcal{F}_{p_U}(k)\rangle^2}.
\]
These fluctuations serve as an estimator of the circuit’s quantum complexity, which we compare to the behavior of a Haar-random ensemble over $N_c N_q$ qubits. The distance to Haar is defined as
\[
D_H = \sqrt{\sum_{k=1}^{2^{N_c N_q}}\Bigl[\mathrm{std}\bigl[\mathcal{F}_{p_U}(k)\bigr] - \mathrm{std}\bigl[\mathcal{F}_H(k)\bigr]\Bigr]^2},
\]
where $\mathcal{F}_H(k)$ are the cumulants corresponding to the Haar ensemble. This metric provides a lower bound reference which, in the large-$N_c N_q$ regime, cannot be reproduced by classical means and allows us to quantify the complexity achieved.

The second approach, semi-analytical in nature, relies on modeling with Markov matrices that describe the statistical evolution of the state moments~\cite{Weinstein}.

Let $\rho$ denote the density matrix of the complete system of $N_c N_q$ qubits. Its expansion in the Pauli basis is
\[
\rho = \frac{1}{2^{N_c N_q}}\sum_{P\in\mathcal{P}}\,r_P\,P,
\]
where $\mathcal{P}$ denotes the set of tensor products of Pauli operators. The average evolution under a random unitary transformation induces a linear transformation on the second moments:
\[
\mathbb{E}\bigl[r_P^2\bigr]\quad\mapsto\quad\sum_Q M_{P Q}\,\mathbb{E}\bigl[r_Q^2\bigr],
\]
where $M$ is a real, stochastic matrix with spectrum in $[0,1]$. This Markov matrix describes how the weight is redistributed among Pauli components in each layer of the circuit.

Following the procedure of Ref.~\cite{Weinstein}, we reduce the full space $\mathbb{R}^{4^{N_c N_q}}$ by statistically identifying the $X$ and $Y$ components of each qubit into a single symbol $\varepsilon$, thereby projecting the dynamics onto the reduced space of symmetric moments of dimension $3^{N_c N_q}$ while preserving the essential convergence properties toward a 2-design.

For single-qubit gates, the local contribution to the reduced matrix is parametrized as
\[
R(c) =
\begin{pmatrix}
1 & 0 & 0 \\
0 & c & \frac{1-c}{2} \\
0 & 1-c & \frac{1+c}{2}
\end{pmatrix},
\]
where $c$ quantifies the degree of randomization. In particular, $c=1/3$ corresponds to fully Haar-random rotations. 

In turn, the two-qubit gates employed in this study are controlled-$Z$ (CZ) operations. Within the Markov matrix formalism of Ref.~\cite{Weinstein}, the action of a CZ gate can be described as a deterministic permutation of the coefficients of the Pauli expansion. In particular, the Markov matrix associated with a CZ gate is a permutation matrix that reorders the components according to the conjugation induced by the gate.

From this construction, we define the Markov matrix of the intracore dynamics as
\[
M_{\mathrm{core}} =
p_1\frac{1}{N_q}\sum_{i=1}^{N_q}\,R^{(i)} \quad+\quad
p_2\frac{1}{N_q(N_q-1)}\sum_{\substack{i,j=1 \\ i\neq j}}^{N_q}\mathrm{CZ}^{(i,j)},
\]
where $R^{(i)}$ denotes the action of a single-qubit gate on qubit $i$, and $\mathrm{CZ}^{(i,j)}$ represents the transformation induced by a CZ gate on the pair $(i,j)$.

The intracore layer corresponds to the tensor product over all cores, describing the dynamics prior to any inter-core connections:
\[
M_{\mathrm{intra}} = \bigotimes_{c=1}^{N_c} M_{\mathrm{core}},
\]
\[
M_{\mathrm{intra}}^{(I)} = \bigl(M_{\mathrm{intra}}\bigr)^{I},
\]
where $I$ denotes the number of consecutive intracore iterations.

The inter-core interaction is modeled by applying CZ gates between qubits belonging to different cores. For each pair $(\alpha,\beta)$ of connected cores, we define
\[
M_{(\alpha,\beta)} =
\frac{1}{2\,N_q^2}\sum_{q_1,q_2=1}^{N_q}\Bigl[
\mathrm{CZ}^{(q_1^{[\alpha]},\,q_2^{[\beta]})}
+
\mathrm{CZ}^{(q_2^{[\beta]},\,q_1^{[\alpha]})}
\Bigr].
\]
The complete inter-core layer is the product over all links allowed by the chosen topology:
\[
M_{\mathrm{inter}} = \prod_{\langle \alpha,\beta\rangle}\,M_{(\alpha,\beta)},
\]
where the product runs over all pairs of cores according to the architecture (linear, ring, star, or fully connected).

Finally, the total Markov matrix describing a full step of the circuit is given by
\[
M_{\mathrm{total}} = M_{\mathrm{inter}}\cdot M_{\mathrm{intra}}^{(I)}.
\]
This matrix acts on the vector of symmetric second moments. Its \emph{spectral gap}, defined as $\Delta=1-\Lambda(I)$, determines the convergence rate toward the stationary regime equivalent to the Haar ensemble in the reduced space. The existence of a maximum of the spectral gap as $I$ is varied constitutes the central result explored in the following sections.

\emph{Numerical simulations of circuit dynamics.}
To illustrate the predictions obtained from the numerical evaluation of our complexity indicators, we performed extensive simulations of variational multicore circuits under various configurations and topologies. In these simulations, we generated ensembles of random circuits with different values of the number of intracore iterations $I$ and characterized the resulting complexity using two independent metrics: the spectral gap and the distance to the Haar measure.

To estimate the spectral gap, we explicitly constructed the Markov matrix corresponding to each combination of the number of cores $N_c$, the number of qubits per core $N_q$, and the chosen interconnection topology. We then diagonalized this matrix and extracted the second-largest eigenvalue $\Lambda(I)$ for $c=1/3$, corresponding to the regime of fully Haar-random single-qubit rotations. Unlike conventional approaches, we defined the normalized spectral gap as
\[
\Delta = 1 - \Lambda^{\,1/D},
\]
where $D$ denotes the total number of gates forming a complete circuit layer:
\[
D = N_c\, I + N_\mathrm{links},
\]
with $N_\mathrm{links}$ being the number of inter-core connections determined by the topology. For example, in the linear architecture $N_\mathrm{links}=N_c-1$, while in the ring $N_\mathrm{links}=N_c$. This definition allows interpreting $\Delta$ as the effective spectral gap per elementary gate, providing an average measure of the convergence rate toward the Haar distribution under the action of a single gate.

In parallel, we quantified operational complexity using the majorization criterion. For each configuration and each value of $I$, we generated an ensemble of 5000 random circuits, computed the average Lorenz curve of the resulting probability distributions, and evaluated the integral of its difference relative to the Lorenz curve of a Haar-random ensemble. We denote this quantity as $ID_H$, which serves as a global metric of the distance to Haar.

\begin{figure}[t]
    \centering
    \includegraphics[width=1.1\linewidth]{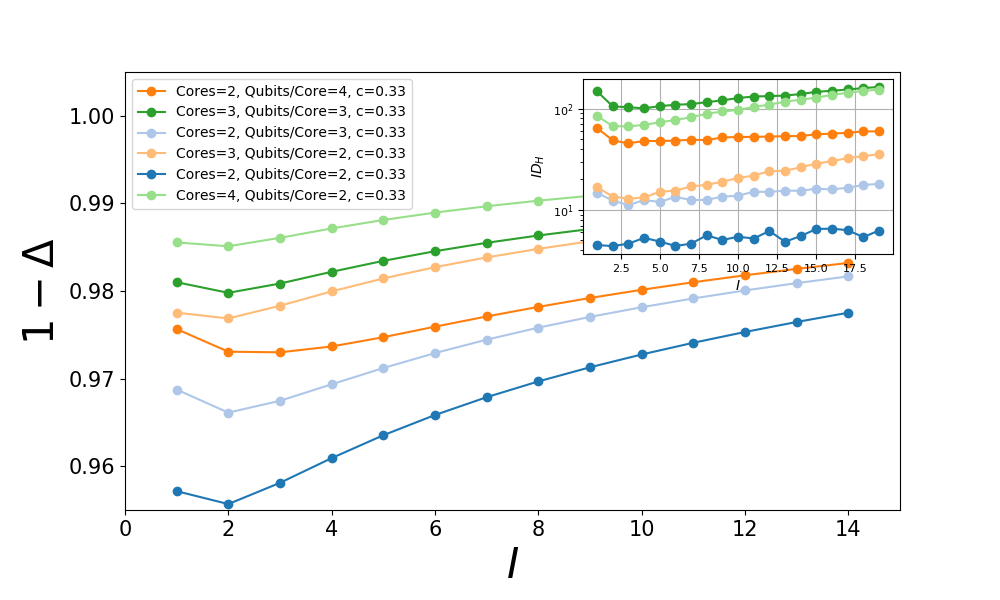}
    \caption{Linear topology. Main panel: $1-\Delta(I)$ as a function of $I$ for different combinations of cores and qubits per core. Inset: integral of the distance to Haar, $ID_H$.}
    \label{fig:fig1}
\end{figure}

Figure~\ref{fig:fig1} shows the results obtained for the linear topology. In the main panel, $1-\Delta(I)$ is plotted as a function of $I$, while the inset displays the evolution of $ID_H$. In all configurations studied, a well-defined minimum emerges, identifying the optimal number of intracore iterations that maximizes the complexity achieved before inter-core operations are applied. For most configurations, the minimum is located around $I\simeq2$, whereas for cores with $N_q=4$ qubits, a slightly higher value near $I\simeq3$ is required. The qualitative agreement between the minima observed in $1-\Delta$ and $ID_H$ confirms that both indicators consistently capture the transition towards the regime of maximal complexity and allow to identify the optimal route. The small discrepancy between the curves arises both from statistical fluctuations and the fact that the majorization measure incorporates information from the entire spectrum of the Markov matrix, whereas $\Delta$ depends only on the second-largest eigenvalue.

\begin{figure}[t]
    \centering
    \includegraphics[width=1.1\linewidth]{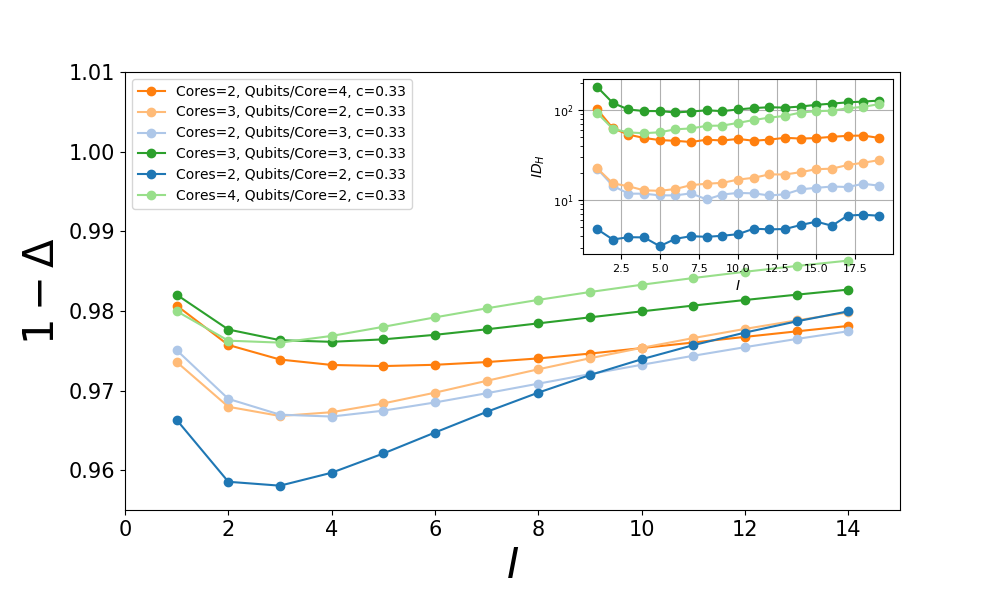}
    \caption{Ring topology. Main panel: $1-\Delta(I)$ as a function of $I$. Inset: $ID_H$.}
    \label{fig:fig2}
\end{figure}

Figure~\ref{fig:fig2} shows the same results for the ring topology. In this case, the agreement between the functional forms of $1-\Delta$ and $ID_H$ is even stronger. As the number of inter-core links increases, entanglement spreads more efficiently across cores, requiring a larger number of intracore iterations to reach maximal complexity. In particular, the minimum shifts toward $I\simeq3$–4, and even close to $I\simeq5$ for configurations with four cores.

\begin{figure}[t]
    \centering
    \includegraphics[width=1.1\linewidth]{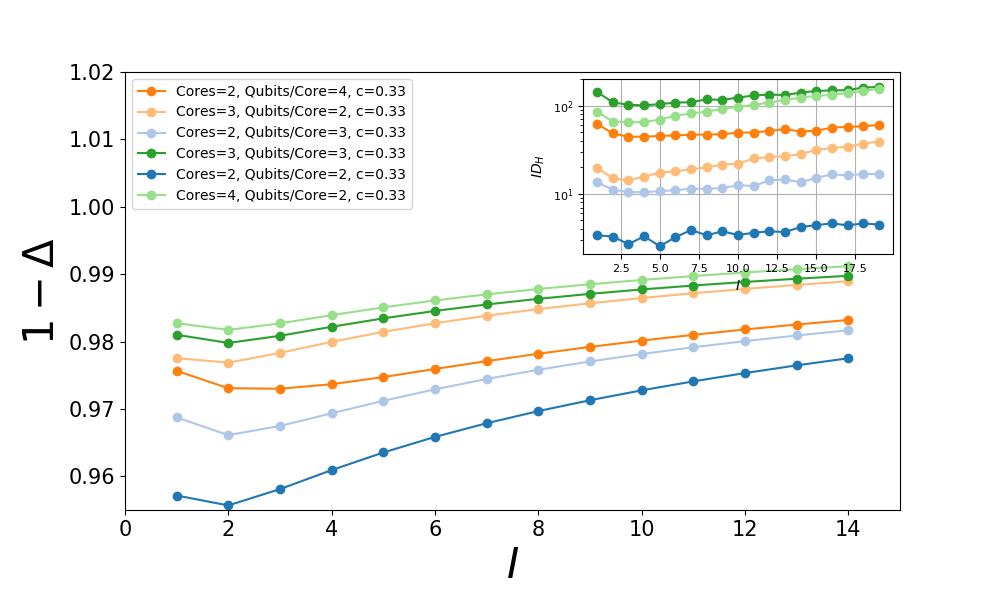}
    \caption{Star topology. Main panel: $1-\Delta(I)$ as a function of $I$. Inset: $ID_H$.}
    \label{fig:fig3}
\end{figure}

Figure~\ref{fig:fig3} displays the results for the star topology. The general behavior is similar to that observed in the linear architecture since, for small numbers of cores, both topologies exhibit comparable connectivity. However, in the case of four cores, a slight reduction in $1-\Delta$ is evident, indicating marginally faster convergence toward the Haar regime due to the higher connectivity of the star configuration.

\begin{figure}[t]
    \centering
    \includegraphics[width=1.1\linewidth]{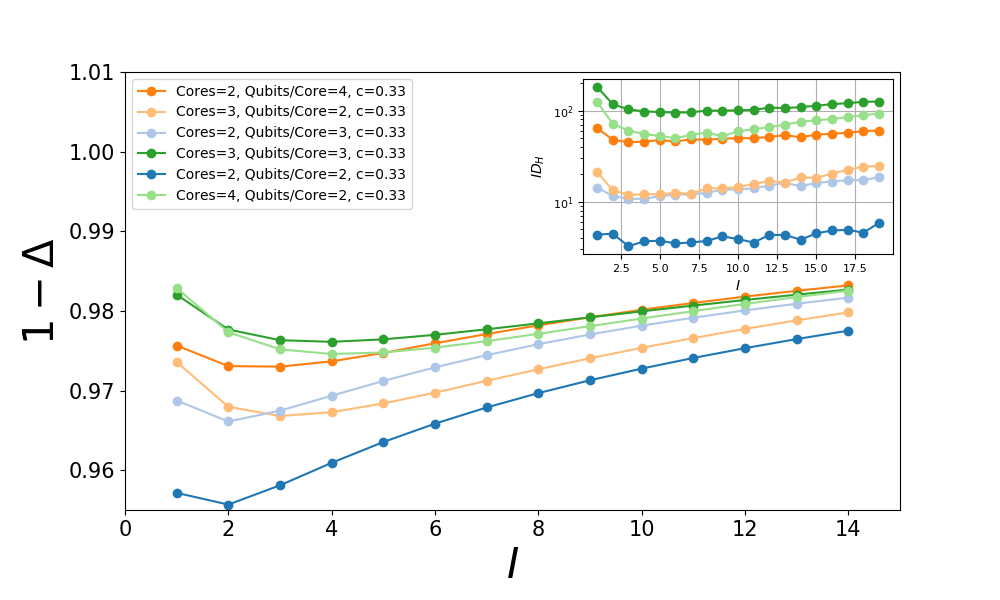}
    \caption{Fully connected topology. Main panel: $1-\Delta(I)$ as a function of $I$. Inset: $ID_H$.}
    \label{fig:fig4}
\end{figure}

Finally, Figure~\ref{fig:fig4} presents the results for the fully connected topology. As expected, the further increase in inter-core links enhances the generation of complexity and shifts the minimum toward higher values of $I$. For four cores, up to $I\simeq5$ intracore iterations are required to achieve maximal complexity.

This set of simulations systematically and robustly confirms the analytical prediction presented below: the existence of an optimal number of intracore iterations that simultaneously maximizes both the effective spectral gap per gate and the operational complexity measured by the distance to Haar. This phenomenon is observed across all topologies and system sizes considered, suggesting that it constitutes a universal property of distributed variational circuits.

\emph{Analytical demonstration of the existence of an optimal configuration.}
In this section, we demonstrate that, for multicore quantum circuits whose Markov matrix describes the evolution in the absence of inter-core gates, the spectral gap $\Delta$ does not exhibit a maximum as a function of the number of intracore iterations $I$. In contrast, when two-qubit gates connecting different cores (e.g., CZ operations) are included, the functional form of the relevant eigenvalues decays more slowly than an exponential, leading to a finite value of $I$ at which the gap reaches a maximum.

We define the spectral gap as
\[
\Delta(I) = 1 - \Lambda(I)^{1/D},
\]
where $\Lambda(I)\in(0,1]$ is a nontrivial eigenvalue of the Markov matrix, strictly below the stationary value $1$.

Our goal is to analyze under what conditions on $\Lambda(I)$ the function $\Delta(I)$ possesses a maximum, i.e., when
\[
\frac{d\Delta}{dI}=0.
\]

For notational convenience, we introduce the parameters $a=N_c$ and $b=N_\mathrm{links}$, so that
\[
\Delta(I)=1-\Lambda(I)^{1/(a I + b)}.
\]
Setting
\[
f(I)=\Lambda(I)^{1/(a I + b)},
\]
we have
\[
\frac{d\Delta}{dI}=-\frac{df}{dI}.
\]
The derivative of $f(I)$ can be computed by applying the chain rule:
\[
\frac{df}{dI}=
\Lambda(I)^{1/(a I + b)}
\left[
-\frac{a \log \Lambda(I)}{(a I + b)^2}
+\frac{1}{a I + b}\cdot\frac{\Lambda'(I)}{\Lambda(I)}
\right].
\]
The critical point where the gap attains a maximum corresponds to the vanishing of this derivative, yielding the condition
\[
(a I + b)\,\frac{\Lambda'(I)}{\Lambda(I)} = a \,\log \Lambda(I),
\]
which constitutes the differential equation that the subleading eigenvalue $\Lambda(I)$ must satisfy for an optimal gate configuration to exist.


By introducing the auxiliary variable
\[
y(I)=\log\Lambda(I),
\]
so that
\[
\frac{\Lambda'(I)}{\Lambda(I)}=y'(I).
\]
The differential equation then reads
\[
y'(I)=\frac{a\,y(I)}{a I + b}.
\]
This is separable and can be integrated explicitly:
\[
\frac{dy}{y}=\frac{a\,dI}{a I + b}\quad\Longrightarrow\quad \log|y|=\log|a I + b|+C,
\]
\[
y(I)=C_1\,(a I + b).
\]
Returning to the original variable,
\[
\log\Lambda(I)=C_1\,(a I + b),
\]
\[
\Lambda(I)=\exp\bigl[C_1\,(a I + b)\bigr].
\]
For $\Lambda(I)$ to decay with $I$, we require $C_1<0$. Setting $C_1=-c$ with $c>0$, we obtain
\[
\Lambda(I)=\exp\bigl[-c\,(a I + b)\bigr]=\lambda\,e^{-c a I},\quad \lambda=e^{-c b}.
\]
This functional form allows a decay slower than a pure exponential in $I$, enabling the existence of a maximum in the gap. As a matter of fact, when the eigenvalue decays more slowly than an exponential, the function $\Delta(I)$ reaches a maximum at a finite number of intracore iterations. This property provides the analytical foundation for the existence of an optimal configuration in multicore quantum circuits with two-qubit gates connecting different cores.

This phenomenon can be understood intuitively by considering how inter-core gates reshape the spectral structure of the Markov matrix. In the absence of inter-core operations, the evolution factorizes into independent dynamics within each core, and the spectrum takes the form
\[
\bigl(\lambda_1^I,\;\lambda_2^I,\;\lambda_3^I,\;\lambda_4^I,\dots\bigr),
\]
where the powers of $I$ reflect the number of intracore iterations. In this regime, each component decays strictly exponentially with $I$. The introduction of two-qubit gates connecting different cores destroys this separability and redistributes entanglement across components. As a result, the decay associated with the second-largest eigenvalue becomes subexponential, since the spectral weight initially concentrated in $\lambda_2^I$ is partially transferred to other modes of the system. This redistribution of spectral contributions is precisely what gives rise to the emergence of a nontrivial maximum in the effective per-gate gap as the number of intracore iterations is varied.

\emph{Conclusion.}
In this work, we have combined analytical derivations, Markov matrix modeling, and extensive numerical simulations to demonstrate that multicore quantum circuits exhibit an optimal number of intracore iterations that maximizes the generation of quantum complexity. In particular, we have shown that the effective spectral gap per elementary gate reaches a nontrivial maximum as a function of the number of intracore layers applied before the inter-core operations, and that this behavior consistently emerges in linear, ring, star, and fully connected architectures.

The detailed analysis of the Markov matrix spectrum, together with the majorization-based metric, enables a quantitative identification of this optimal regime and provides a clear interpretation of its origin: the balance between local randomization within each core and the global entanglement induced by inter-core connections. Our results establish that this balance universally yields maximal operational complexity in distributed variational circuits.

These findings offer a general principle that can guide the design of scalable quantum processors and supply a quantitative criterion to assess the capability of multicore architectures to efficiently approximate Haar-random ensembles. 

Respect to future research, it will be of great interest to explore the impact of noise and gate errors, as well as to extend this approach to other classes of entangling gates.

\section{Acknowledgments}
This work has been partially supported by the Spanish Ministry of Science, 
Innovation and Universities, Gobierno de Espa\~na, under Contract No.\ PID2021-122711NB-C21. Support from CONICET is greatfully acknowledged.

\begin{acknowledgments}

\end{acknowledgments}

\bibliographystyle{apsrev4-2}
\bibliography{references}

\end{document}